\documentstyle[preprint,aps,floats,epsf,fleqn]{revtex}
\newcommand{\ds}{\displaystyle}
\newcommand{\ee}{{\bf e}}
\newcommand{\rr}{{\bf r}}
\newcommand{\sss}{{\bf s}}
\begin{document}
\newfont{\eufxii}{eufm10 scaled 1200}
\newfont{\eusix}{eusm10 scaled 900}
\newfont{\eusxii}{eusm10 scaled 1200}
\newfont{\eurxii}{eurm10 scaled 1200}
\draft
\setcounter{page}{0}
\title{
CONTINUUM MONTE CARLO SIMULATION
AT CONSTANT PRESSURE 
OF STIFF CHAIN MOLECULES AT SURFACES}
\author{F.M. Haa$\mbox{\rm s}^1$ and R. Hilfe$\mbox{\rm r}^{1,2}$}
\address{
$\mbox{ }^1$Institut f{\"u}r Physik,
Universit{\"a}t Mainz,
55099 Mainz,
Germany\\
$\mbox{ }^2$Institute of Physics, University of Oslo,
P.O. Box 1048, 0316 Oslo, Norway\\
{\rm present address}: ICA-1, Universit{\"a}t Stuttgart, 
70569 Stuttgart, Germany}
\maketitle
\thispagestyle{empty}
\begin{abstract}
Continuum Monte-Carlo simulations at constant
pressure are performed on short chain molecules
at surfaces.
The rodlike chains, consisting of seven effective monomers, 
are attached at one end to a flat twodimensional substrate.
It is found that the model exhibits phases similar
to the liquid condensed and liquid expanded phases 
of Langmuir monolayers.
The model is investigated
here for a wide range of pressures and temperatures
using a special form of constant pressure simulation
compatible with the symmetry breaking during tilting
transitions in the liquid condensed phases.
At low pressures the chains undergo a tilting transition
exhibiting tilt directions towards nearest and also next 
nearest neighbours depending on temperature.
At elevated temperatures and low pressure the film enters
a fluidlike phase similar to
the liquid expanded phase observed in experiment.
\end{abstract}
\newpage

\section{Introduction}
Amphiphilic molecules, such as fatty acids 
($CH_3(CH_2)_nCOOH$) or phospholipids,
form monomolecular layers on water surfaces 
in which the hydrophobic headgroups are 
immersed into water while the hydrophobic
alkane chains point outwards \cite{BK37}.
This phenomenon is utilized in 
the Langmuir-Blodgett coating technique, in which
under suitable conditions on $pH$ and temperature,
monolayer or multilayer films can be transferred to 
a solid substrate by successively dipping the substrate 
into a compressed layer on a water surface.
Langmuir-Blodgett films are highly ordered and their molecular
ordering depends on the structure of the compressed layer on
the water surface prior to dipping \cite{WV89}.
The ordering of Langmuir-Blodgett films plays an important 
role in applications ranging from microlithography
and electrooptics to biochemical sensing and tribology.

Many experiments have in recent years 
investigated the molecular structure and phase behaviour
of Langmuir monolayers (see \cite{WV89,moe90,KD92,ABKR94} 
for reviews).
These studies were spurred by the possibility of
combining modern x-ray diffraction methods with  
more traditional thermodynamic measurements \cite{BKP91,KBBPMAK91}. 
The rodlike non axially symmetric molecules give rise to
a large variety of phases and phase transitions \cite{KD92}
with a rich polymorphism.

Despite differences in detail monolayers of fatty acids, 
phopholipids, alcohols and esters exhibit 
a similar phase diagram \cite{BP90,BKP91,ABKR94}.
This fact alone suggests that the structural similarities between 
amphiphilic molecules are responsible for the similarities
observed in the phase diagrams.
Hence the chemical and atomistic differences between different
molecules may not be necessary to understand the phase diagram
of amphiphiles.
This observation motivates our studies of highly idealized coarse 
grained models to elucidate the mechanisms underlying various transitions.
Among these transitions the tilting transition in the liquid
condensed phases has been our main interest.
We study idealized models in order to separate the influence of various 
coarse parameters (such as chain flexibility, symmetry of the 
molecules, mobility and geometry of head groups etc.) 
from that of finer details (such as interaction potentials,
cis-trans conformations, hydrogen atoms etc.)
on the tilting transition in these monolayers.

Given the advances in numerical simulation techniques
there have already been numerous simulation studies of chemically realistic 
\cite{HR88,EB88,BCK88,HK89,BK90,MTKQ91,BC91,KTO92,CSR92,SCR92,ABS93}
and idealized coarse grained models 
\cite{SRG86,KKB90,MKS90,hil92c,hil93a,HLB93} 
for the tilting transition in amphiphilic monolayers.
Chemically realistic models with much atomistic detail have an
advantage in their ability to describe dense packing effects more
realistically.
Molecular dynamics simulations of such models are very useful
for elucidating the behaviour of such models away from phase
transitions. 
Near a phase transition Monte Carlo methods can be more 
efficient for obtaining the statistical averages.
Highly idealized coarse grained models may be simpler (and
sometimes faster) to simulate but they often suffer from neglecting
some aspects of the physics. 
For example, treating the alkane chains as rigid cylindrical rods,
originally suggested in \cite{SRG86}, disregards the melting transition
of the layer, and the intrachain conformational disorder.
Such rigid rods grafted onto a triangular lattice were studied 
extensively in \cite{KKB90,hil92c,hil93a} using continuum 
Monte Carlo calculations.
While this model exhibits a tilting transition it cannot
show a liquid expanded phase or the restructuring and melting 
of the head group lattice.
Although the grafting of the head groups in this model severely 
limits the applicability of the model to Langmuir layers 
the model is useful and realistic for chains that are 
permanently grafted to a solid substrate \cite{WV89}.
Chemically realistic as well as coarse grained models suffer 
from the general limitation that the effective potentials are 
only inaccurately known \cite{bin95}.

Layers of flexible chain molecules with mobile head groups
\cite{hil94f,hil94h} are intermediate between chemically 
realistic united atom models and idealized coarse grained models 
such as rigid rods or even lattice models \cite{HLB93}.
Such models 
were studied in \cite{hil94f,hil94h} using continuum Monte 
Carlo simulations at fixed volume.
It was found that the 
condition of fixed volume imposes a symmetry breaking field 
which can suppress the restructuring of the head group lattice
during the tilting transition.
This was concluded from the presence of metastable boundary induced
defect structures \cite{HHB95}.
While such defects are known to exist in experiment and are therefore 
of interest in their own right \cite{FBK94}, they make the extraction of
thermodynamic equilibrium properties from small scale
simulations very difficult.
Hence it is of great interest to perform also simulations at 
constant pressures in which the symmetry breaking field is absent.

In the present paper we report simulations at constant pressure
using an algorithm which is compatible with the symmetry
breaking during tilting transitions.
To facilitate comparison we investigate 
a model similar to the one studied at constant area
in \cite{hil94f,hil94h}.
Our main objective is twofold: 
Firstly we would like to confirm the existence of an expected 
orthorhombic distortion of the head group lattice during a tilting
transition in a coarse grained model.
Such observations are essential for the interpretation 
of constant volume simulations in which the distortion is suppressed.
For example, the impossibility to distort or contract the simulation
box in such simulations combined with periodic boundary conditions 
can lead to the appearance of negative pressures at low densities 
if the interaction potential has an attractive part.
Secondly, we want to understand the selection mechanisms for nearest 
versus next nearest neighbour tilt direction and the rather high 
transition temperature of the first order melting transition observed 
in previous constant volume simulations of the same model
\cite{hil94f,hil94h,haa95,HHB95}.
In \cite{hil94f,hil94h} the fluidized phase appeared at
temperatures roughly four times higher than the tilting 
transition temperature.
This seems to rule out the possibility of identifying 
the molten state with the liquid expanded phase of
Langmuir monolayers.
An important result of the present study is to show that 
in constant pressure simulations the fluidized phase appears
already at much lower temperatures.

\section{Model description}

Each amphiphilic molecule is represented through a 
chain of seven effective monomers labeled 
$i=0,...,6$ with $i=0$ corresponding to the 
hydrophilic head group of the amphiphile. 
The cartesian coordinate 
system in threedimensional space is chosen such that 
the headgroup $i=0$ is restricted to move in the $z=0$
plane representing the twodimensional substrate.
All the effective monomers are connected through a 
finitely extendable nonlinear elastic (FENE) potential
\begin{equation}
V_{bl}(d) = \left\{
\begin{array}{r@{\quad:\quad}l}
-\ds\frac{c_{bl}d_{bl}^2}{2}
\ln\left(1-\ds\frac{(d-d_0)^2}{d_{bl}^2}\right)
& \mbox{for}\quad |d-d_0| < d_{bl} \\[12pt]
\infty & \mbox{for}\quad |d-d_0|\geq d_{bl}
\end{array}
\right .
\end{equation}
where $d$ is the distance between monomers, and 
$c_{bl}>0$ is the spring constant. 
The FENE potential is harmonic at its minimum $d_0$
but the bonds cannot be stretched beyond a maximum
length determined by $d_{bl}$.
The stiffness of the rodlike molecules whose alkane 
chains are predominantly in an all trans conformation is 
incorporated by a bond angle potential
\begin{equation}
V_{ba}(\theta_i) = c_{ba}(1+\cos(\theta_i))
\end{equation}
where $c_{ba}$ is the force constant and $\theta_i$
is the angle formed by the three monomers $i-1,i,i+1$.
Note that $V_{ba}$ is a three body interaction.
All monomers except nearest neighbours within the 
same chain interact through a Lennard-Jones potential.
The Lennard-Jones potential is truncated and shifted such 
that it vanishes at the truncation point. 
If $\epsilon$ is the interaction strength and $\sigma$
its range then
\begin{equation}
V_{LJ}(d) = \left\{
\begin{array}{r@{\quad:\quad}l}
\epsilon((\sigma/d)^{12}-2(\sigma/d)^6-(1/d_{LJ})^{12}+2(1/d_{LJ})^6)
& \mbox{for}\quad d \leq d_{LJ}\sigma \\[12pt]
0 & \mbox{for}\quad d > d_{LJ}\sigma
\end{array}
\right .
\end{equation}
where $d=d_{LJ}\sigma$ is the truncation point.

The model above is similar to the model used
for the constant volume simulations in \cite{hil94f,hil94h}.
It extends and generalizes an earlier 
study of a system consisting of perfectly rigid rods 
grafted to a hexagonal lattice and interacting with
Lennard-Jones interactions \cite{hil92c,hil93a}.
The main difference with \cite{hil94f,hil94h}
consists in simulating at constant
spreading pressure rather than constant substrate
area, and in replacing the cutoff harmonic potential
for the bondlengths with a FENE potential.

The earlier simulations at constant volume were
found to lead to boundary induced defects \cite{HHB95}.
These can significantly perturb the
long range correlations at phase transitions.
Most molecular dynamics simulations are similarly carried 
out at constant volume or aspect ratio of the simulation box.
Simulations at constant pressure on the other
hand reproduce the experimental situation more
exactly, and allow the defects to relax via 
distortions of the head group lattice.

\section{Relation between the model and experiment}
\label{relation}

The model presented in the previous section is not intended
to be a chemically realistic model for Langmuir monolayers.
On the contrary we wish to study a highly idealized coarse 
grained model in order to understand the influence of varying
degrees of idealization on the mechanisms responsible for the
tilting transitions in the condensed phases.
The model has resulted from previous systematic 
investigations of even more idealized models
\cite{KKB90,hil92c,hil93a,HLB93,hil94f,hil94h}.

Although the model is not intended to be chemically realistic 
it is of interest to connect the model with 
reality through a rough mapping of length and energy scales.
Such a rough mapping can be helpful in deciding 
whether the phase transitions of the model may be related to 
the phase transitions of the experimental system.

If a typical fatty acid with chain lengths from 12 to 30 
carbons is represented by a model chain with seven effective
monomers then each effective monomer of our model chains 
must represent roughly between two and five methyl groups.
Hence, in such a mapping the diameter of the effective monomers
is of order 10\AA, and the Lennard-Jones interaction range $\sigma$ 
is of order 5\AA. 
For the energy scale a value for $\epsilon$ of order 200K 
is a very rough estimate.

The simple attachment of the head groups to the 
substrate represents an important overidealization
of our model.
Experimentally the form and interactions of the
head groups appear to play an important role.
While our model is more realistic than fully 
discrete lattice models \cite{HLB93} it contains 
less chemical detail than united atom models which 
have been investigated using molecular dynamics 
simulations \cite{HR88,BK90,BC91}.
The main difference is that our molecules are
axially symmetric while the zigzag structure
of united atom models destroys this symmetry.
Consequently we do not expect the model to
reproduce phases associated with a herringbone 
ordering.
We feel, however, that our model while neglecting
these features represents a good compromise between 
computational efficiency and chemical realism.

\section{Simulation details}

Lengths and energies in our simulations will be
measured in dimensionless units defined by 
setting $\sigma=1$ and $\epsilon=1$. In these
units the other parameters for the simulations 
are chosen as $d_0=0.7,d_{bl}=0.2,d_{LJ}=2.0$ and 
$c_{bl}=100, c_{ba}=10$ which ensures that the
chains do not intersect.
The choice $c_{bl},c_{ba}\gg \epsilon$ 
produces stiff rodlike chains.
The system, consisting of 144 chain molecules, is 
confined to an area of sidelengths $L_x,L_y$ in the 
$z=0$ plane.
The height of the simulation box $L_z\gg 6d_0$ is chosen 
much larger than the length of the chain molecules.

A simplified groundstate analysis of the model was
performed previously \cite{hil94f,hil94h,haa95}.
It predicts a tilting transition between a tilted
and an untilted state, and highlights the importance
of packing effects.

The continuum model is simulated in a canonical ensemble 
in which temperature, spreading pressure and particle number
are kept constant.
The simulation is carried out using a Metropolis Monte Carlo
procedure in which individual monomer positions and the area
$A=L_xL_y$ of the simulation box are updated in continuous space. 
The continuous position space requires 
to use methods adapted from molecular dynamics simulations
for evaluating the interaction energies. 
The maximal jump distance of monomers in a single move is
chosen to optimize the acceptance rate while ensuring that
the chains cannot intersect each other.
The cutoff in the interaction potentials allows us to use
an adaptation of the link cell algorithm to keep track 
of interacting neighbouring monomers.
The simulation box is subdivided into cells of sidelength
larger than or equal to the interaction radius.
Particles in a subcell can only interact with particles 
in neighbouring cells. 
The cell contents are stored using specially 
designed linked pointer lists.
The algorithm was developed and tested for constant
volume simulations \cite{hilxx,hil94f,hil94h},
and its speed depends strongly on the density
and the interaction ranges.

The partition function $Z_{N\Pi_A T}$ of the isobaric ensemble 
is obtained from the isochoric partition function $Z_{NAT}$
through Laplace transformation. Denoting the surface area by $A$
and the spreading pressure by $\Pi_A$ it reads
\begin{equation}
Z_{N\Pi_A T} = \int_0^\infty\int\exp\left(-\beta(\Pi_A A + V(\rr))\right)\;d\rr\;dA
\end{equation}
where $\beta=1/(k_B T)$, $k_B$ is the Boltzmann constant,  $d\rr$ is the 
integral of all coordinates, and all constant prefactors have been suppressed.
Expectation values for an observable $X(\rr)$ in this ensemble are 
calculated as
\begin{equation}
\langle X \rangle = \frac{1}{Z_{N\Pi_A T}}
\int_0^\infty\int X(\sss) \exp\left(-\beta(V(\sss)+\Pi_A A)+N\ln A\right)
\;d\sss\;dA
\end{equation}
in terms of the scaled coordinates $\sss=(r_x/L_x,r_y/L_y,r_z)$
of all the particles.
The expectation values can be calculated with the usual Metropolis
scheme by introducing an effective Hamiltonian
\begin{equation}
H_{eff}(\sss,L_x,L_y) = V(\sss) + \Pi_A L_xL_y - Nk_BT\ln(L_xL_y)
\end{equation}
containing two additional degrees of freedom $L_x$ and $L_y$.
Attempts to change the area $A=L_xL_y$ of the simulation 
box are carried out after each Monte Carlo step 
(1MCS = 1 update/monomer) of regular coordinate updates.
A maximal step size of $\Delta L_x=\Delta L_y=\pm 0.01$ for the
area moves was found to yield optimal acceptance rates for the
chosen system size.
The evaluation of the energy difference $\Delta V(\sss)$ for
an area move involves a recalculation of all particle interaction
potentials because the coordinates are rescaled differently in
the $x$- and $y$-directions.
This makes area moves computationally expensive and leads to
a deceleration of the constant pressure algorithm by a factor
of roughly two compared with constant area simulations.

The simulations are started from an untilted configuration 
of $12\times 12$ chain molecules. 
The chains are positioned on a triangular lattice and directed 
perpendicular to the substrate.
If a tilted initial configuration with hexagonal headgroup
lattice is used at low temperatures the system was found to
evolve into metastable states with lifetimes well beyond 
$10^4$ Monte Carlo steps.
In all simulations periodic boundary conditions are 
applied in the $xy$-directions.
In each run the system was first equilibrated for
20000 Monte Carlo steps (updates per monomer). Subsequently
averages were recorded every 500th MCS over a period of
50000 MCS. 
These calculations consumed several 100 hours
of CPU time on IBM RS6000 370 equipment.

The twodimensional pressure tensor was estimated form the
forces on each particle using the virial theorem and the
minimum image convention appropriate for periodic boundary
conditions \cite{AT87}.
While the contribution from the bondlength and Lennard-Jones
interactions to the pressure can be obtained as usual, the 
contribution from the three body bond angle interaction 
$V_{ba}$ has to be considered separately. 
It is found that the contribution of the bond angle interaction
$V_{ba}$ can be neglected.

\section{Results}

At low temperatures, $T=0.2$ the sidelengths $L_x,L_y$ fluctuate 
very little, and the system organizes itself into a configuration 
consisting of 12 rows and 12 columns of chain molecules.
At $T=2.0$ the sidelengths fluctuate more strongly, and these
fluctuations are anticorrelated in time.
Increases in $L_x$ are accompanied by decreasing $L_y$
and vice versa yielding roughly constant area $A$.
In general the measured pressure agrees well with the 
applied pressure. 
Small deviations are attributed to finite size effects,
and the fact that the pressure is measured from the
virial theorem rather than from the free energy.

Figure \ref{configs} shows snapshots of equilibrated
\marginpar{\fbox{\em Fig. \ref{configs}}} 
configurations of 144 chains at temperatures $T=0.2,1.0,2.0$ 
and pressures $\Pi_A=10,100$.
Each effective monomer is represented as a sphere of
radius $\sigma/2$.
At high pressures $\Pi_A=100$ the chains are 
compressed into a hexagonal arrangement in which all
chains stand perpendicular to the substrate.
At the low temperature $T=0.2$ and high pressure the
monolayer surface shows a nearly regular modulation
resulting from extension and contraction of the
chains along the molecular axis.
At higher temperatures the surface is rough.
At low pressures the chains assume a uniformly tilted
crystalline arrangement which progressively disorders 
as the temperature is increased.
At $T=2.0$ the layer has molten into a fluidlike
phase.

The crystalline order and the melting of the surface
layer can be seen from Figure \ref{dichte} showing
\marginpar{\fbox{\em Fig. \ref{dichte}}} 
the density profiles along the $z$ direction.
Figure \ref{dichte} shows also results at an
intermediate pressure $\Pi_A=30$.
At low temperature $T=0.2$ the system is crystalline.
For $\Pi_A=10$ (and $T=0.2$) there are six pronounced
maxima in the density profile.
The position of last maximum corresponding to the 
tail group $i=6$ is significantly shifted from the
value $6 d_0 = 4.2$ corresponding to untilted
chains. This indicates that the chains are tilted.
At higher pressures $\Pi_A=30$ and $\Pi_A=100$ (and low
temperatures $T=0.2$) the density profile exhibits
$11$ respectively $12$ maxima instead of $6$.
This fact together with the positions of the
maxima shows that the chains are alternatingly
stretched and contracted along the molecular
axis. 
The concomitant doubling of the number of monomer layers 
has also been observed in simulations at constant
volume when the density becomes as high as 1.3 \cite{haa95,HHB95}.
At the intermediate temperature $T=1.0$ and $\Pi_A=10$ the
crystalline order is stabilized by the tilt order.
For $\Pi_A=30$ and $\Pi_A=100$ the surface layer
is begininning to melt.
At $T=2.0$ the tail group layer ($i=6$) is molten for all
pressures but the layered structure is still
intact in the lower layers.
Note however that for $T=2.0$ and $\Pi_A=10$
the second layer ($i=1$) is spreading into the substrate
surface ($z=0$) indicating a melting of the head group
lattice.

Defining the monolayer thickness from the largest
(rightmost) inflection point along the density profile 
allows to analyze the change in monolayer thickness as 
a function of temperature and pressure.
The results are collected in Table \ref{thickness}.
As expected the thickness increases with increasing 
temperature and with increasing spreading pressure.
\begin{table}\caption{\label{thickness}\small
Monolayer thickness for different temperatures and
pressures}
\vspace*{24pt}
\begin{tabular}{lccc}
 & $\Pi_A=10$ & $\Pi_A=30$ & $\Pi_A=100$ \\ \hline\hline
$T=0.2$ & 3.7 & 4.0 & 4.1 \rule[-12pt]{0pt}{32pt} \\ \hline
$T=1.0$ & 3.9 & 4.1 & 4.2 \rule[-12pt]{0pt}{32pt}\\ \hline
$T=2.0$ & 3.9 & 4.2 & 4.3 \rule[-12pt]{0pt}{32pt}
\end{tabular}
\end{table}

To analyze the tilt order we display in Figure \ref{voronoi}
\marginpar{\fbox{\em Fig. \ref{voronoi}}} 
the Voronoi diagrams for the configurations shown above in
Figure \ref{configs}.
In these figures each chain molecule is shown as a dot
with a short line attached to it.
The dot represents the center of the head group, and
the short line represents the projection of the 
head to tail vector into the substrate plane.
In addition the Voronoi cells are drawn around each
head group to visualize the degree of crystallinity
in the head group lattice.
These figures show that at high pressures $\Pi_A=100$ the
chain molecules are untilted at all temperatures,
although defects and randomly oriented projections
of small magnitude are visible at elevated temperatures.
At $\Pi_A=100$ the head group lattice is crystalline at all 
temperatures.
At low pressure and low temperature ($\Pi_A=10,T=0.2$)
the figure shows tilting in agreement with the density
profiles.
The tilt is directed towards the nearest neighbour molecule.
At $\Pi_A=10,T=1.0$, however, the average tilt direction
is towards next nearest neighbours.
Both of these tilt directions
have been observed in experiment \cite{KD92}.
At high temperatures ($\Pi_A=10,T=2.0$) both the tilt
order and the crystalline ordering of the head
groups are destroyed.

The expected magnitude of the absolute value of the tilt angle 
$\langle|\theta|\rangle$, defined as the 
angle between the end-end vector and the surface normal,
is displayed over a wide pressure range in Figure \ref{tilt}
\marginpar{\fbox{\em Fig. \ref{tilt}}}
for the temperatures $T=0.2,1.0,2.0$.
At all temperatures the tilt angle increases as the
spreading pressure is lowered.
However, such an increase does not necessarily indicate
a larger collective tilt \cite{hil92c,hil93a,hil94f,hil94h}.
An increase in tilt magnitude can also be caused by an 
increase in the fluctuations of the tilt angle.
The two situations can be distinguished by measuring 
additional quantities.
The projection $R_{xy}$ of the end-end vector $\ee=(e_x,e_y,e_z)$
into the $xy$-plane is defined as
\begin{equation}
R_{xy}=\sqrt{\langle(\overline{e_x}^2+\overline{e_y}^2)\rangle}
\end{equation}
where $\overline{e_x}$ and $\overline{e_y}$ are the configuration 
average of the $x$ and $y$ components of the end-end vector $\ee$.
The quantity $R_{xy}$ is independent of the tilt direction.
Figure \ref{order}, which shows $R_{xy}$ over the same pressure
\marginpar{\fbox{\em Fig. \ref{order}}}
range as the tilt angle, indicates that the jump in
$\langle|\theta|\rangle$ at low pressures has different origins
for $T=0.2$ and $T=2.0$.
For $T=0.2$ the simultaneous increase in $R_{xy}$ indicates
a tilting transition at around $\Pi_A\approx 30$.
The smooth behaviour of $R_{xy}$ for $T=2.0$ indicates
that the jump in $\langle|\theta|\rangle$ at this temperature
is not caused by tilting but by melting.
This interpretation is suggested also by the analysis
of the configurations and density profiles shown in
Figures \ref{configs}--\ref{voronoi}.
It is further corroborated by measuring the orientational
correlations between neighbouring chains.
The orientational correlations are measured through the quantity
\begin{equation}
K_{NN}=\left\langle
\frac{1}{6}\sum\frac{1}{2}(3\cos^2(\theta_{NN}-1)
\right\rangle
\end{equation}
where $\theta_{NN}$ is the angle between the end-end vectors
of two nearest neighbour chains, and the sum extends over all
nearest neighbour chains of a given chain.
The nearest neighbour chains are defined as those six chains 
whose head groups have the smallest distances from the given
chain.
Because of steric hindrances the correlation $K_{NN}$ is
usually very strong.
Figure \ref{correl} shows the pressure dependence of $K_{NN}$
\marginpar{\fbox{\em Fig. \ref{correl}}}
at $T=2.0$ which exhibits a sharp drop at low pressures.
The same curves for $T=0.2$ and $T=1.0$ are indistinguishable
from the line at $K_{NN}=1$.
The sharp drop in Figure \ref{correl} indicates the presence 
of a melting transition as suggested by the jump in
$\langle|\theta|\rangle$ at the same pressures.

The melting transition at $T=2.0$ and low pressures appears
also when plotting the specific area $L_xL_y/N$ ($N=144$) 
or the ratio of sidelengths $L_x/L_y$ as a function of 
pressure as shown in Figure \ref{distort}.
\marginpar{\fbox{\em Fig. \ref{distort}}}
The area per molecule shows a steady increase from values
around 0.72 to 0.93 and then a sudden jump to a value
close to 1.2.

At low temperatures $T=0.2$ the specific area shows a sudden
increase at the critical pressure $\Pi_A\approx 30$ for the
tilting transition.
This increase is associated with a similar increase in the ratio 
$L_x/L_y$ reflecting an orthorhombic distortion of the head 
group lattice during the tilting transition.
At high pressures the ratio $L_x/L_y$ approaches the value
$2/\sqrt{3}$ for the hexagonal lattice independent of
temperature.

\section{Discussion}

Our present simulation results combined with previous 
simulations at constant area \cite{hil94f,hil94h,haa95,HHB95} 
indicate that the idealized coarse grained model has 
at least four distinct phases: an ordered phase without
tilt, two ordered phases with tilt towards nearest and next
nearest neighbours respectively, and a highly disordered
fluidized phase.

At sufficiently high pressures the chains organize into
a hexagonally ordered condensed phase without tilt.
This phase persists throughout the whole investigated
temperature range, and it
can be understood as a close packing configuration.
As the pressure is increased the individual monomers 
begin to move away from the minima in the bond angle
potentials, and this leads to stretching and contraction
of the chains along their molecular axis resulting in
the visible periodic modulation of the free surface.
The modulation of the surface is expected to depend 
on the relative strength of the bond length potential 
as compared to the Lennard-Jones and the bond angle 
potential.

At lower pressures the chains tilt either in the direction
of the nearest neighbour chain, or in the direction of the
next nearest neighbour chain.
The tilting may be understood as a result of the Lennard-Jones 
attraction between neighbouring chains.
The tilt direction is the result of a complicated interplay
between packing, energetic and entropic effects.
At spreading pressure $\Pi_A=10$ and temperature $T=0.2$
we observe nearest neighbour tilt combined with a periodically
modulated stretching of the chains along their axis.
This indicates that packing effects are dominant for this
pressure.
At higher temperatures, $T=1.0$, the tilt direction changes
into the energetically favoured next nearest neighbour direction.
A complete understanding of these effects will require not
only more simulations at intermediate pressures and temperatures
but in addition a systematic exploration of the space of
interaction parameters which is beyond the scope of the 
present study.

Finally at still higher temperatures the individual chain 
conformation and the head group lattice become disordered. 
The system enters a fluidized phase in which
more and more configurations become accessible 
as the temperature is increased.

A similar sequence of phases has been observed also
in simulations at constant area when varying temperature
and density \cite{hil94f,hil94h,haa95,HHB95}.
Contrary to those simulations the phases appear here in a much
narrower temperature range.
In particular the fluidized phase appears already at much lower
temperatures than in simulations at constant area.
If the rough estimate $\sigma\approx 5$\AA from section
\ref{relation} is used to map the model to experiment 
then the area per molecule increases from values around 
23\AA/molecule to 30\AA/per molecule as seen in
Figure \ref{distort}.
This agrees roughly with the experimental values for the 
transition from liquid condensed to expanded phases, and
suggests to identify the fluidized phase of our model with 
the liquid expanded phase in Langmuir layers.

Because of the wide range of pressures studied here
the pressure resolution is poor, and hence our results 
do not allow to identify the order of this transition 
from the present study.
Finite size scaling analyses with higher pressure and 
temperature resolution are needed to conclusively 
identify the order of this transition.

Our simulation at constant pressure using an algorithm 
compatible with the symmetry breaking allows to answer
the questions, posed in the introduction, concerning
the restructuring of the head group lattice during tilting.
Figure \ref{distort} demonstrates the existence of an 
orthorhombic distortion of the head group lattice during 
the tilting transition.
The magnitude of this effect may also be estimated from
Figure \ref{distort}.
This distortion is suppressed in simulations at constant 
area because the simulation box is rigid in that case.
A second effect of the rigid simulation box in constant
area simulations with periodic boundary conditions is
the appearance of negative pressures at low densities
and temperatures \cite{haa95}.
The negative pressure arises from the attractive part of the
Lennard-Jones interactions and complicates the analysis
of simulations at constant area.
Our constant pressure simulations presented here are
free from such complications.

In summary we find that our coarse grained model
exhibits untilted, tilted and fluidized phases similar
to those observed in experiment.
The model reproduces the positional ordering,
bond orientational ordering and tilt ordering
similar to that observed in the condensed phases 
of the experiment \cite{BKP91,KBBPMAK91,KD92}.
It does not allow for herringbone ordering, however,
because of the cylindrical symmetry of the chains.
Our simulations at constant pressure show for the first
time the experimentally known orthorhombic distortion of 
the head group lattice during tilting transitions in the
liquid condensed phases.
We have also shown that the choice of ensemble 
(constant pressure versus constant area) plays an 
important role in small scale simulations as 
evidenced by the large difference in transition 
temperatures for the appearance of the fluidized phase.
This will become particularly important when trying to
make quantitative predictions.

Of course it is clear that an idealized model such as
the one studied in this paper can only be a small step
in understanding the complexity of Langmuir monolayers,
and improvements of the model are desirable in several
directions. 
In particular the important role played by the head
groups and their interactions are inadequately 
represented in our model.
Similarly the cylindrical symmetry of the chains
is unrealistic, as mentioned previously.
Improving the model and simulation technique step
by step, and studying the intermediate idealized
models, it will ultimately become possible to 
understand quantitatively the full complexity of the 
phase structure observed in the experiment.
\newpage
ACKNOWLEDGEMENT: We are grateful to Prof.Dr. K. Binder for
discussions, and we thank
the Deutsche Forschungsgemeinschaft through 
its Graduiertenkolleg ``Physik und Chemie supramolekularer Systeme''
(F.M.H.) and Norges Forskningsrad (R.H.) for financial support.

\begin{figure}[p]
\caption{
Snapshots of equilibrated configurations containing 144 molecules
at various temperatures and spreading pressures corresponding to
a) $T=0.2,\Pi_A=10$;
b) $T=0.2,\Pi_A=100$;
c) $T=1.0,\Pi_A=10$;
d) $T=1.0,\Pi_A=100$;
e) $T=2.0,\Pi_A=10$;
f) $T=2.0,\Pi_A=100$.
}
\label{configs}
\vspace*{24pt}
\end{figure}
\begin{figure}[p]
\caption{
Total monomer density profiles as a function of the distance $z$
from the substrate. The densities are normalized to 6. The head 
group monomer ($i=0$) at $z=0$ is not included in the plot.
}
\label{dichte}
\vspace*{24pt}
\end{figure}
\begin{figure}[p]
\caption{
Voronoi diagrams of equilibrated configurations shown
in Figure \protect\ref{configs} at various temperatures 
spreading pressures corresponding to 
a) $T=0.2,\Pi_A=10$;
b) $T=0.2,\Pi_A=100$;
c) $T=1.0,\Pi_A=10$;
d) $T=1.0,\Pi_A=100$;
e) $T=2.0,\Pi_A=10$;
f) $T=2.0,\Pi_A=100$.
The corresponding box dimensions are
$L_x(T=0.2,\Pi_A=10)=11.87$,
$L_y(T=0.2,\Pi_A=10)=9.82$,
$L_x(T=0.2,\Pi_A=100)=10.92$,
$L_y(T=0.2,\Pi_A=100)=9.45$,
$L_x(T=1.0,\Pi_A=10)=11.58$,
$L_y(T=1.0,\Pi_A=10)=10.21$,
$L_x(T=1.0,\Pi_A=100)=11.15$,
$L_y(T=1.0,\Pi_A=100)=9.63$,
$L_x(T=2.0,\Pi_A=10)=14.49$,
$L_y(T=2.0,\Pi_A=10)=11.61$,
$L_x(T=2.0,\Pi_A=100)=11.38$ and
$L_y(T=2.0,\Pi_A=100)=9.93$.
}
\label{voronoi}
\vspace*{24pt}
\end{figure}
\begin{figure}[p]
\caption{
Average tilt angle $\langle|\theta|\rangle$ as function of
spreading pressure for temperatures $T=0.2,1.0,2.0$
}
\label{tilt}
\vspace*{24pt}
\end{figure}
\begin{figure}[p]
\caption{
Averaged projection of the end-end vector $R_{xy}$ into the substrate 
plane as function of spreading pressure for temperatures $T=0.2,1.0,2.0$
}
\label{order}
\vspace*{24pt}
\end{figure}
\begin{figure}[p]
\caption{
Orientational correlation between neighbouring chains 
as function of spreading pressure for $T=2.0$
}
\label{correl}
\vspace*{24pt}
\end{figure}
\begin{figure}[p]
\caption{
Averaged specific area $L_xL_y/N$ and sidelength ratio $L_x/L_y$
as function of spreading pressure for $T=0.2,1.0,2.0$
}
\label{distort}
\end{figure}

\bibliographystyle{prsty}
\bibliography{$HOME/tex/bib/books,$HOME/tex/bib/langmuir,$HOME/tex/bib/proceedings,$HOME/tex/bib/reviews,$HOME/tex/bib/publ}
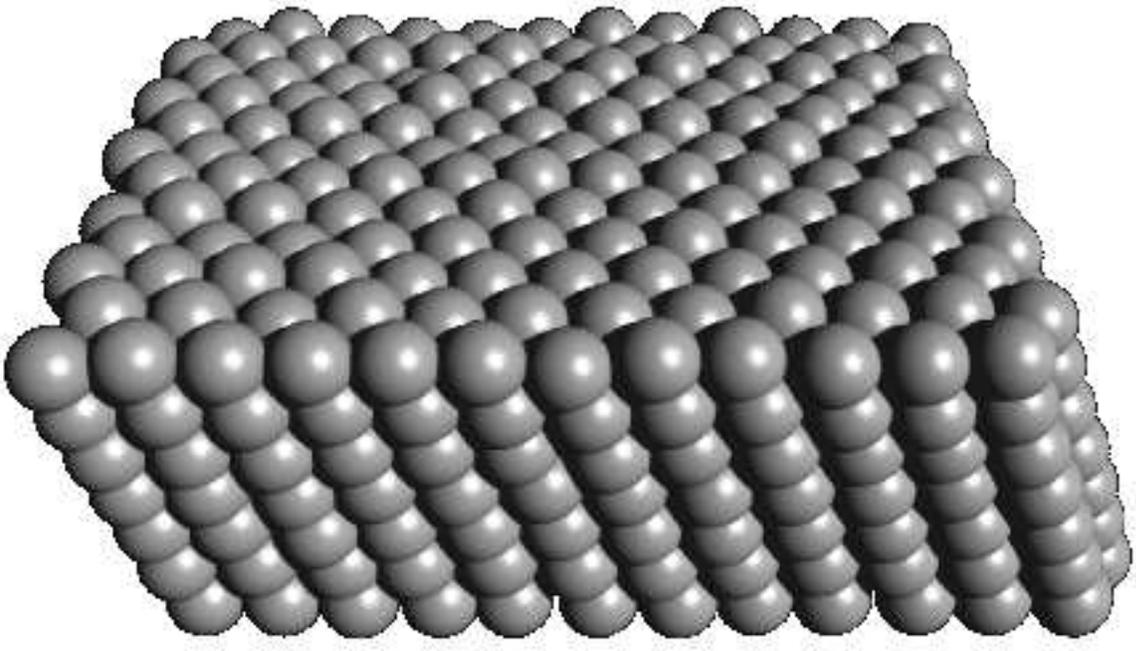
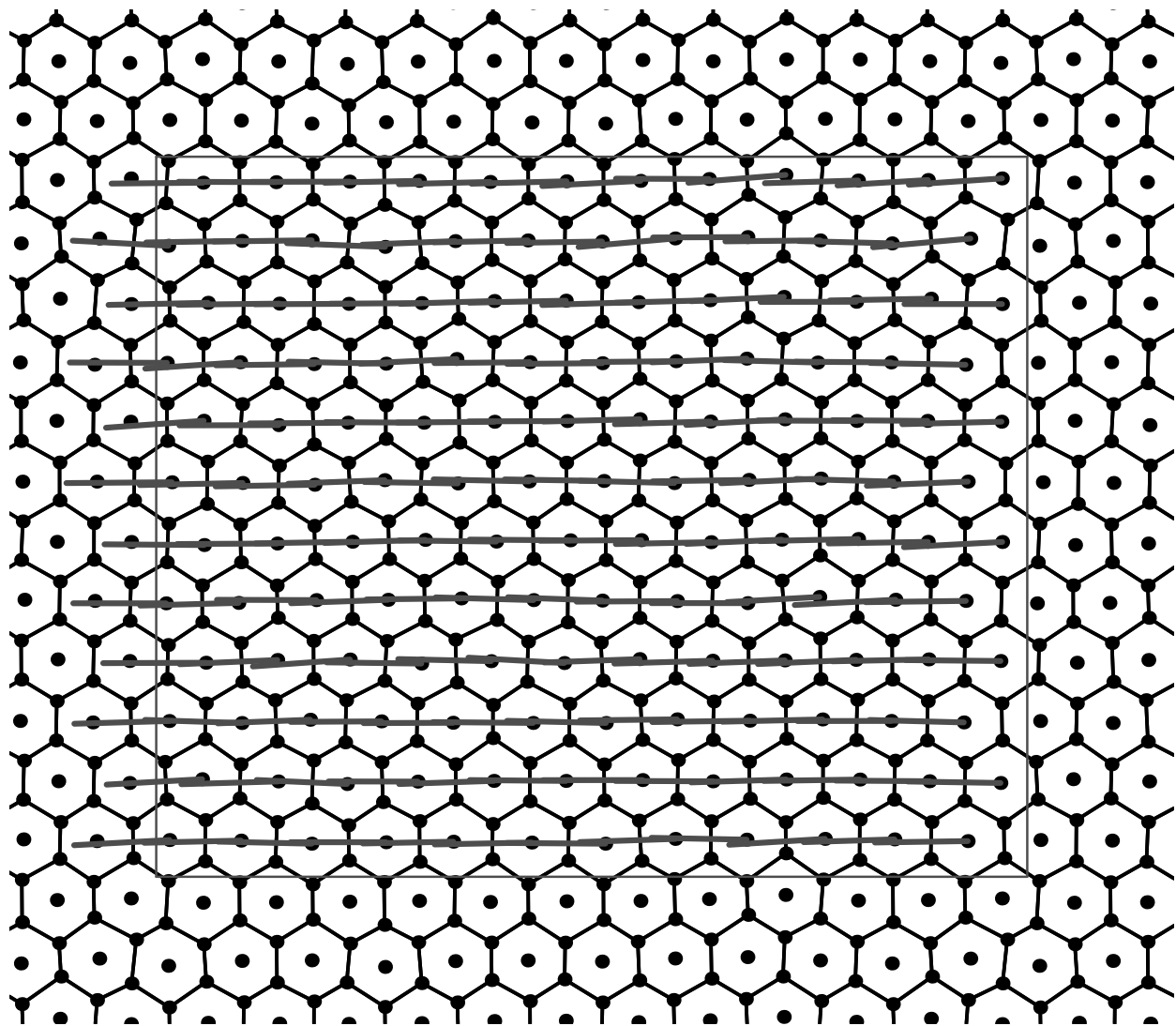
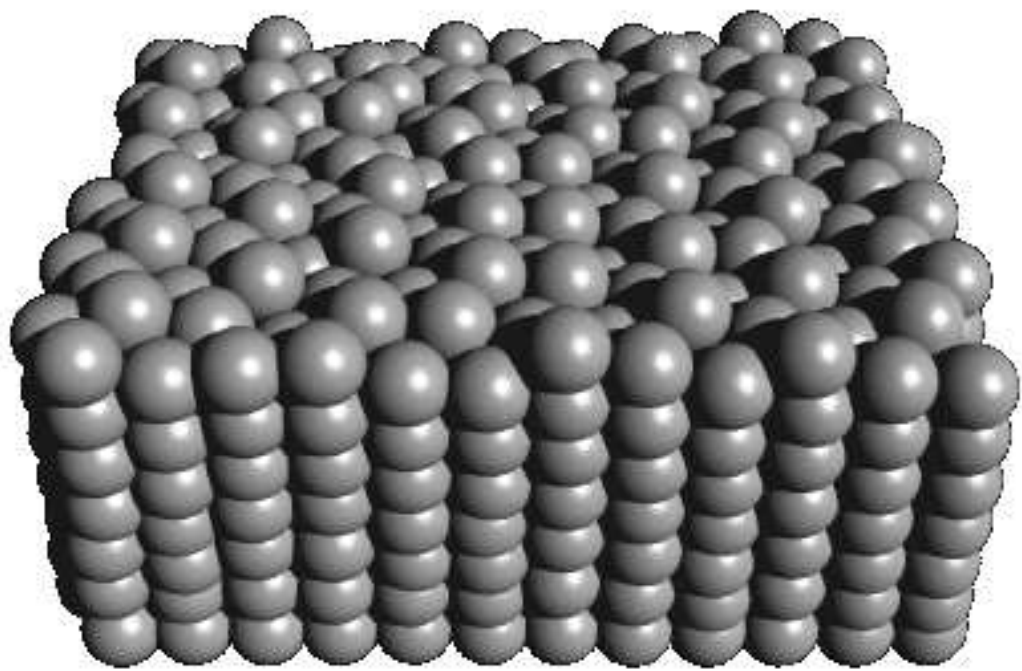
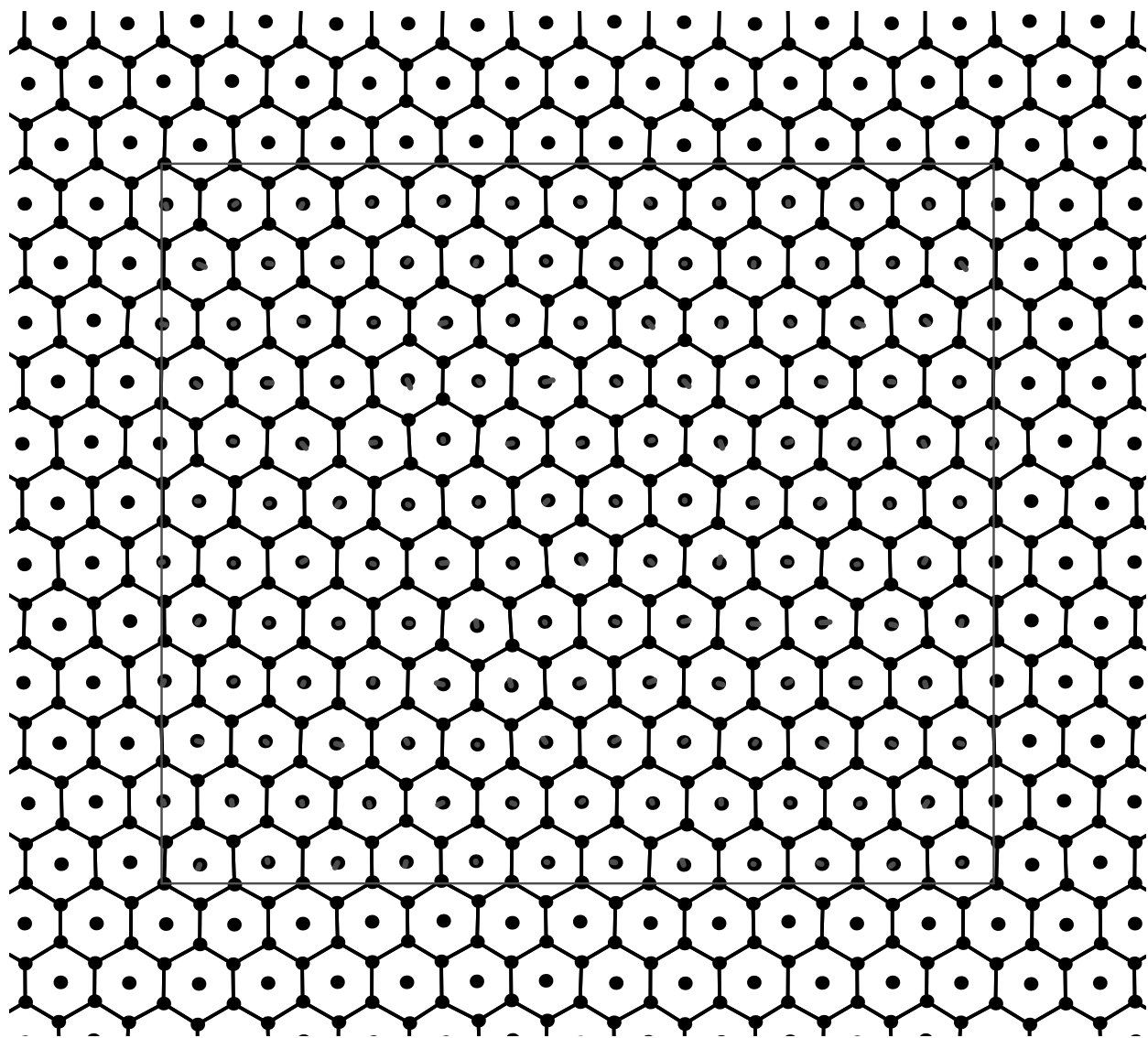
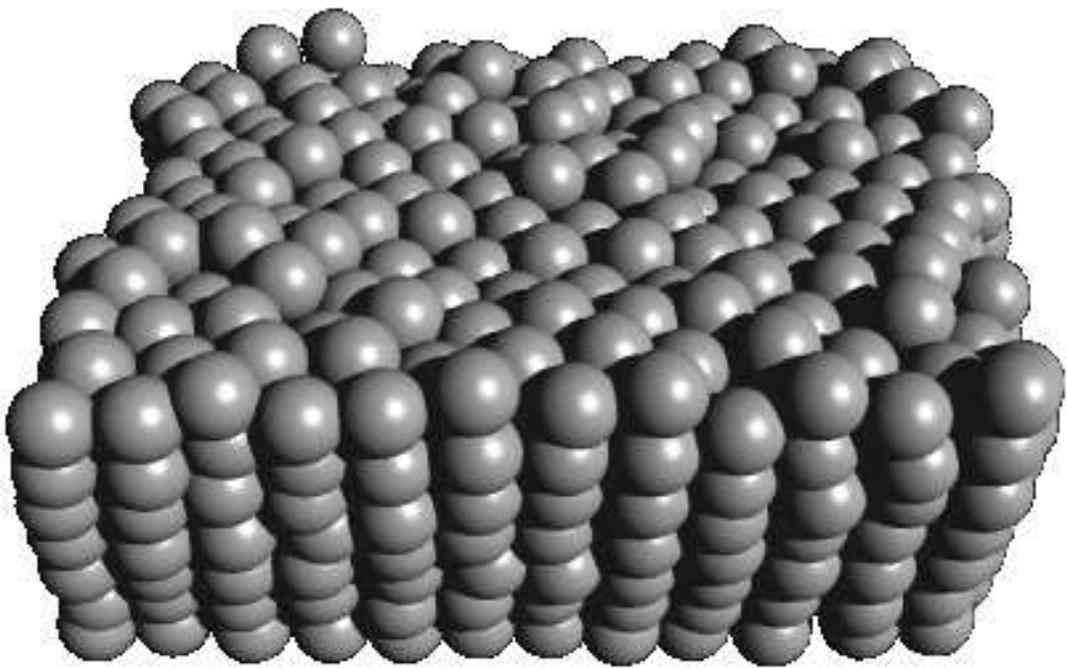
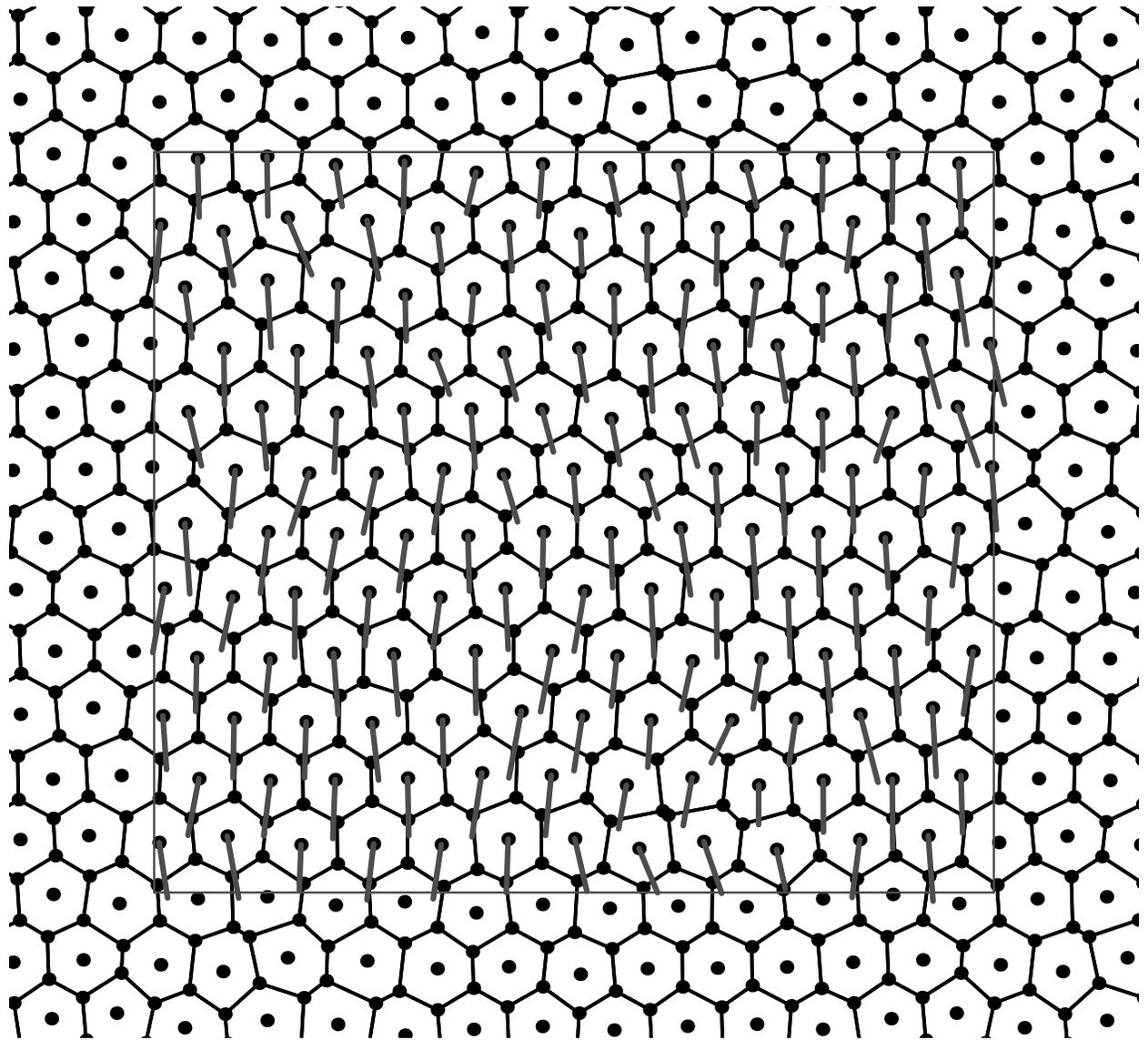
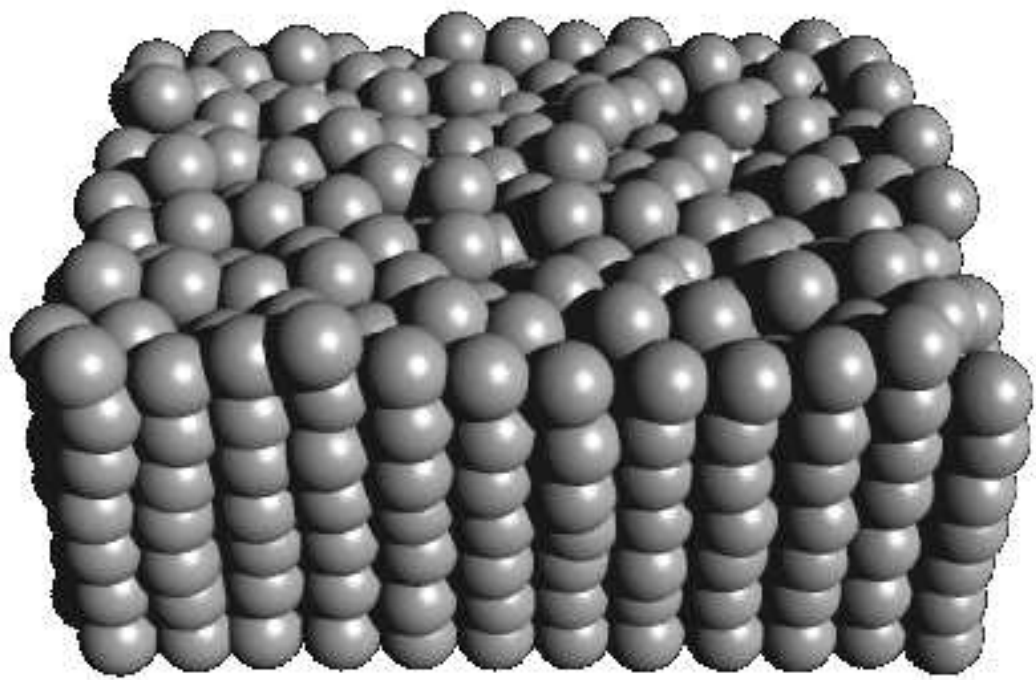
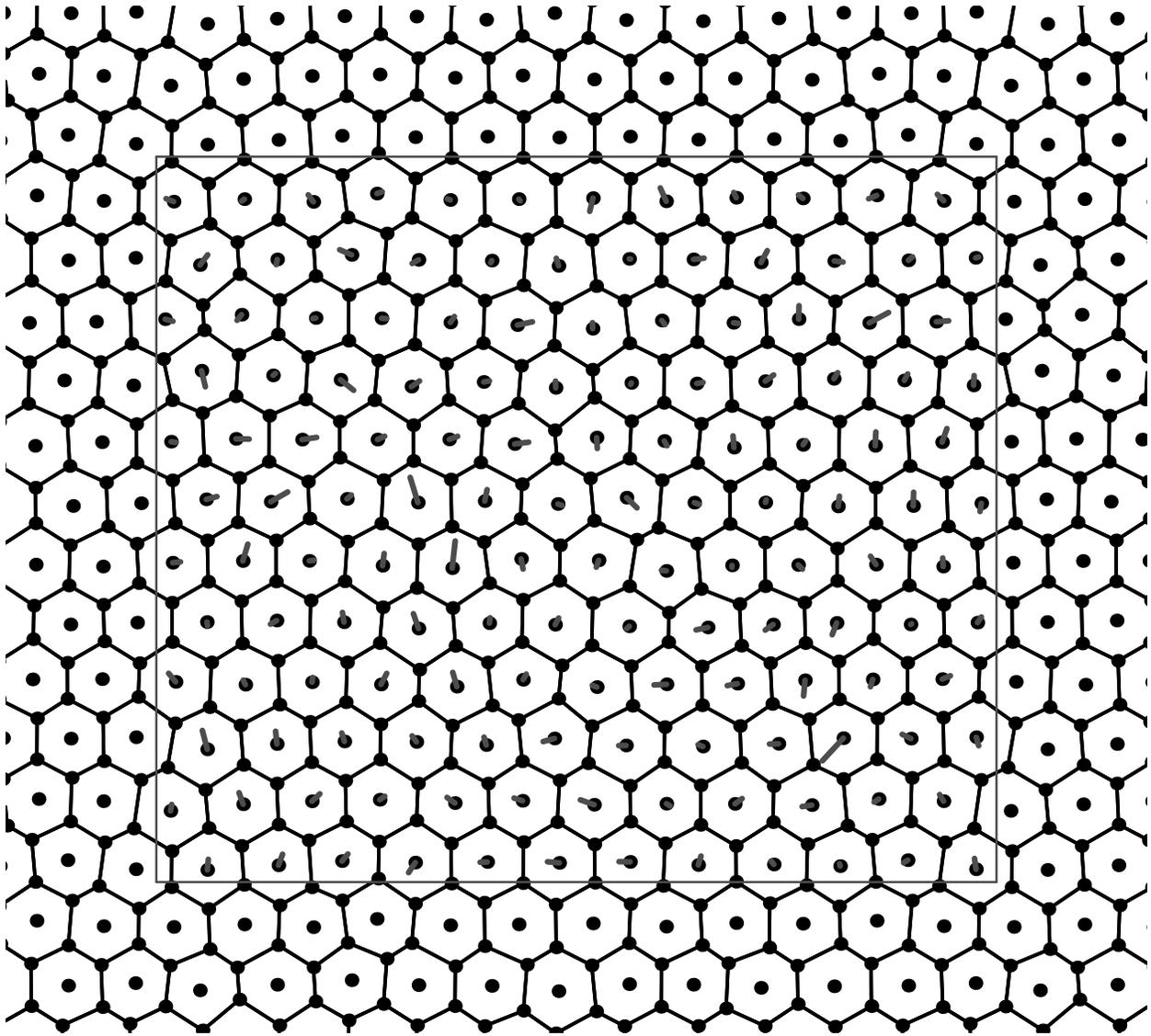
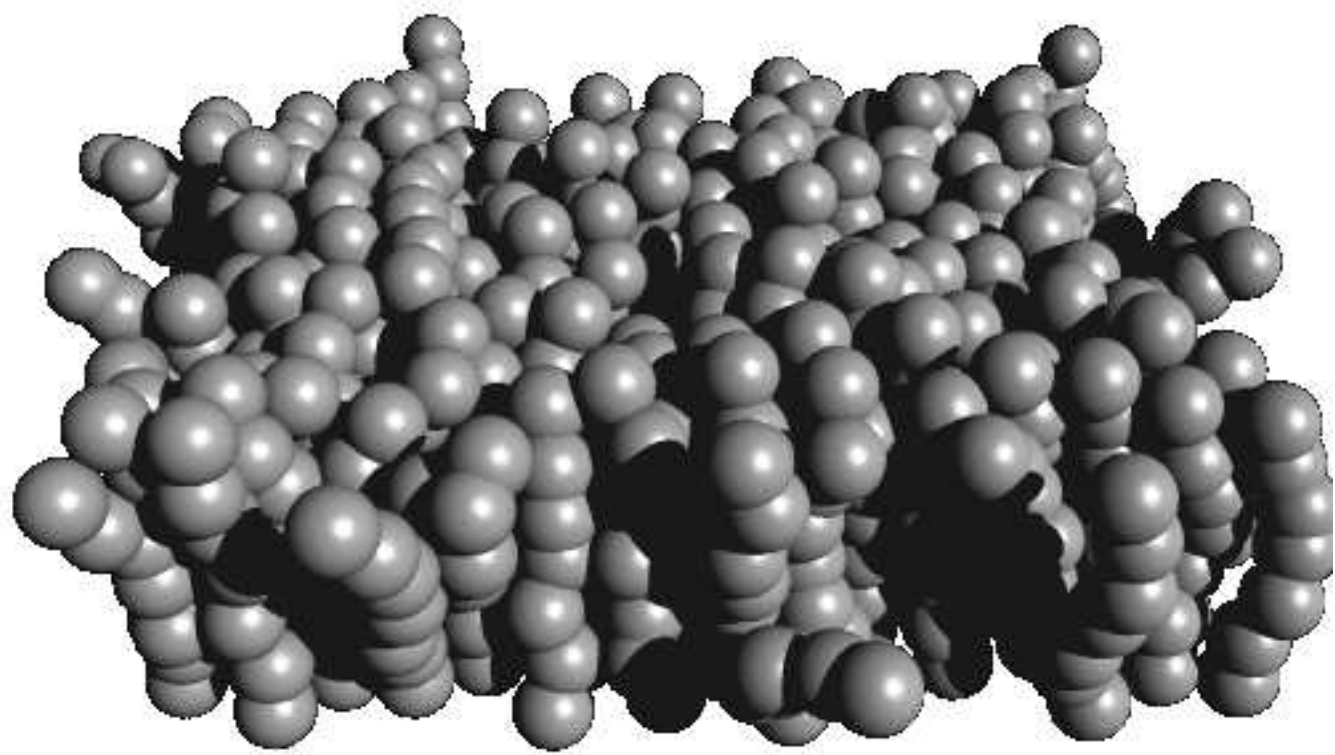
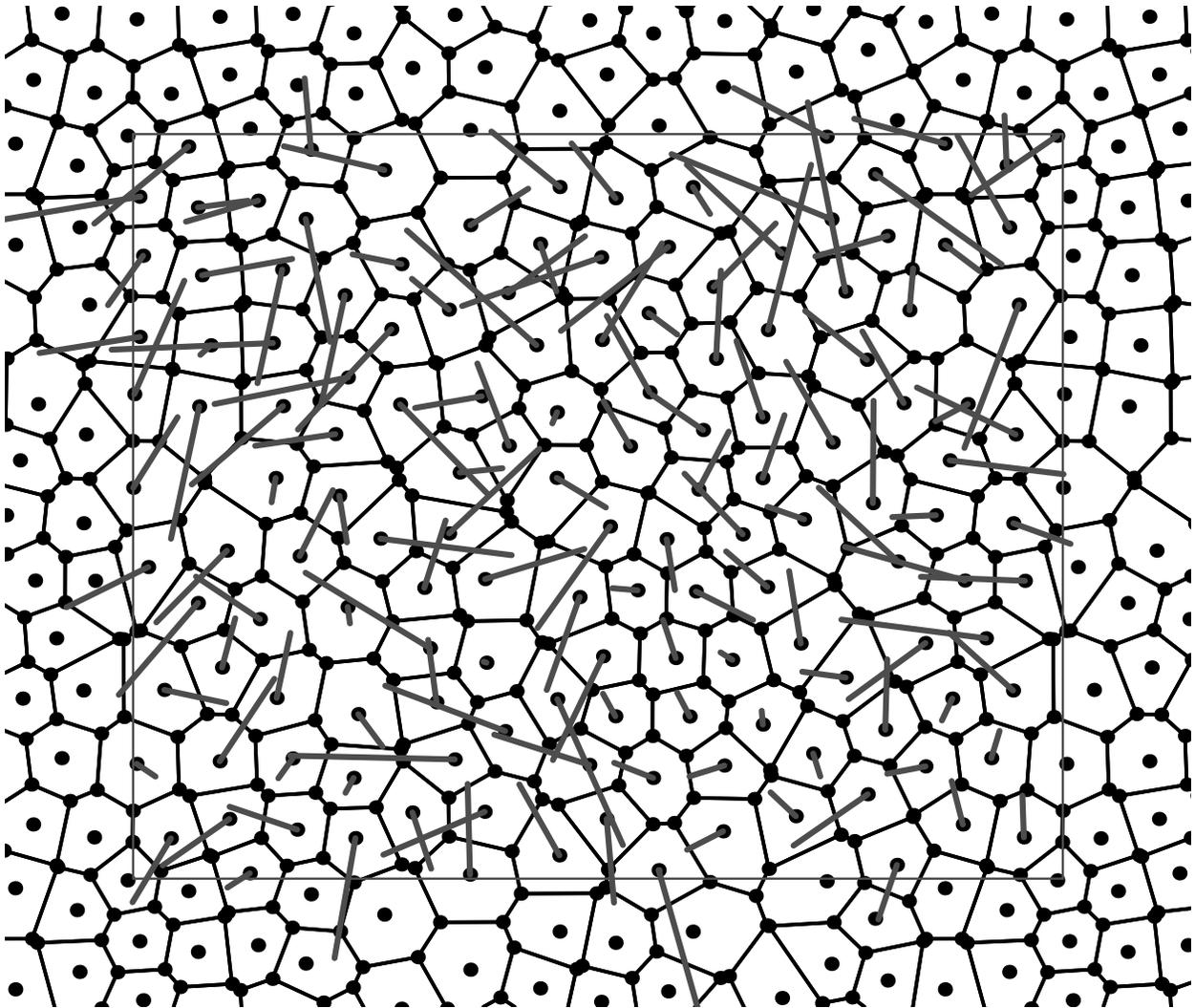
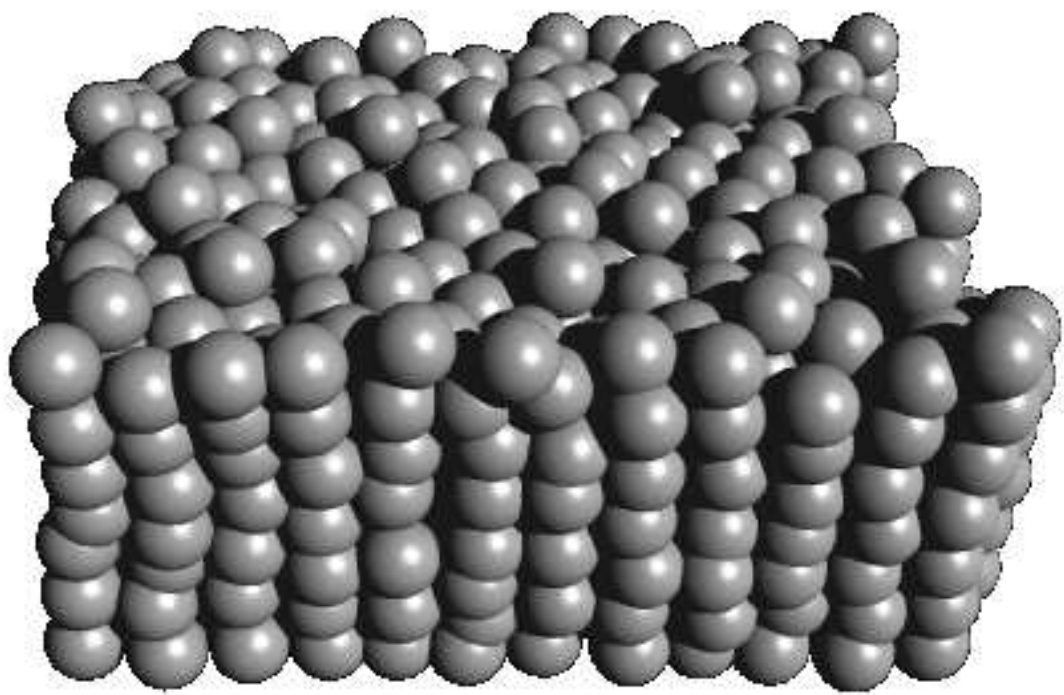
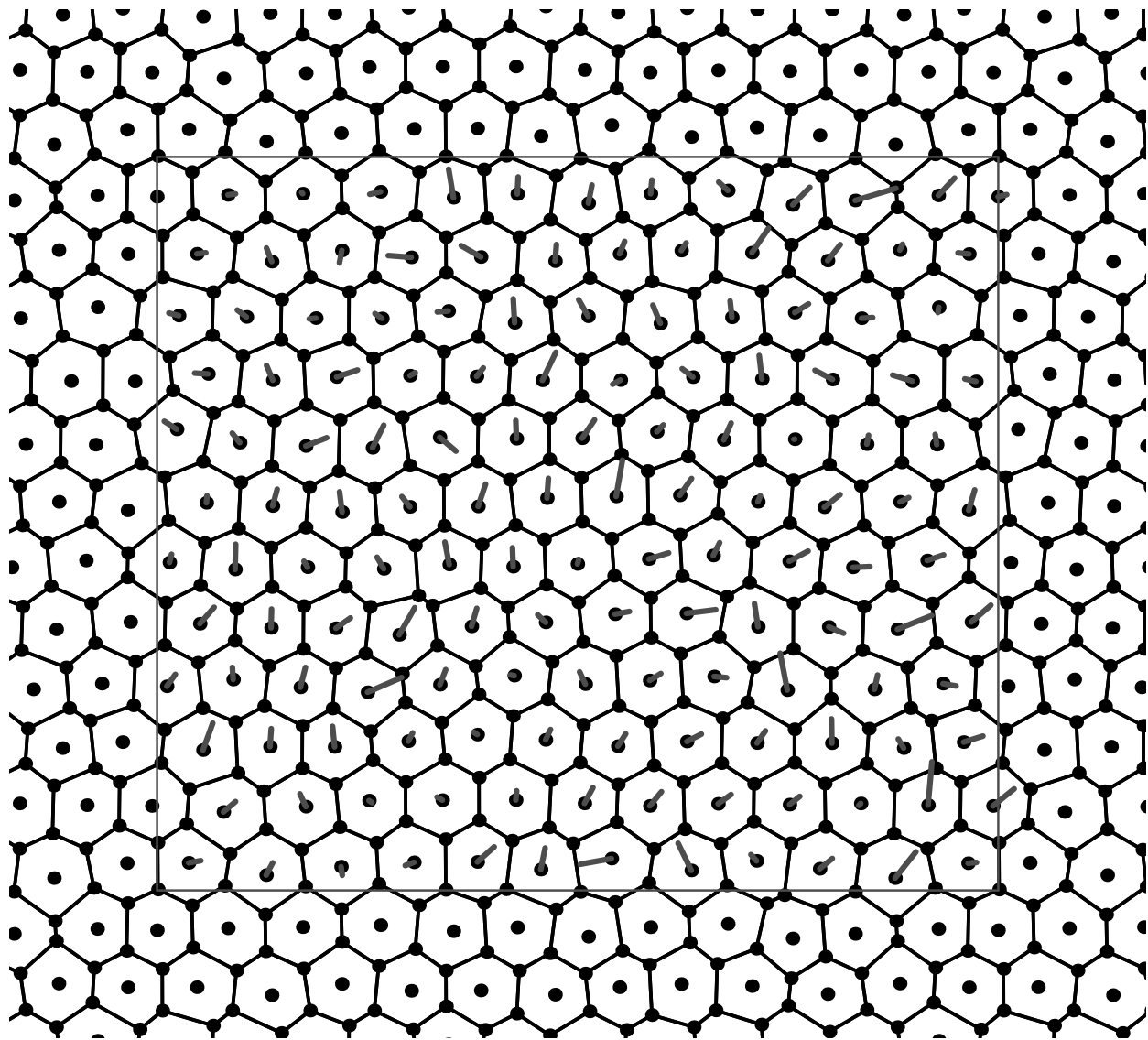
\end{document}